\documentclass[superscriptaddress, twocolumn]{revtex4-1}
\usepackage{amsmath, amssymb, amstext}
\usepackage{graphicx}
\usepackage[T1]{fontenc}% Use modern font encodings
\usepackage{rotating}
\usepackage{multirow}
\usepackage{adjustbox}
\usepackage{chemformula}
\usepackage{hyperref}
\usepackage{enumitem}

\begin{document}

\title{Quantum-chemistry-aided identification, synthesis and experimental validation of model systems for conformationally controlled reaction studies: Separation of the conformers of 2,3-dibromobuta-1,3-diene in the gas phase}

\author{Ardita Kilaj}
\affiliation{Department of Chemistry, University of Basel, 4056 Basel, Switzerland}
\author{Hong Gao}
\affiliation{Beijing National Laboratory for Molecular Sciences, Institute of Chemistry, Chinese Academy of Sciences, Beijing 100190, China}
\author{Diana Tahchieva}
\affiliation{Department of Chemistry, University of Basel, 4056 Basel, Switzerland}
\author{Raghunathan Ramakrishnan}
\affiliation{Centre for Interdisciplinary Sciences, Tata Institute of Fundamental Research, Hyderabad 500107, India}

\author{Daniel Bachmann}
\affiliation{Department of Chemistry, University of Basel, 4056 Basel, Switzerland}
\author{Dennis Gillingham}
\affiliation{Department of Chemistry, University of Basel, 4056 Basel, Switzerland}
\author{O.\ Anatole von Lilienfeld}
\affiliation{Department of Chemistry, University of Basel, 4056 Basel, Switzerland}
\affiliation{National Center for Computational Design and Discovery of Novel Materials (MARVEL), University of Basel, 4056 Basel, Switzerland}

\author{Jochen~K\"upper}

\affiliation{Department of Physics, Universit\"at Hamburg, 22761~Hamburg, Germany}
\affiliation{Department of Chemistry, Universit\"at Hamburg, 20146~Hamburg, Germany}
\affiliation{Center for Free-Electron Laser Science, Deutsches Elektronen-Synchrotron DESY, 22607~Hamburg, Germany}
\affiliation{Center for Ultrafast Imaging, Universit\"at Hamburg, 22761~Hamburg, Germany}

\author{Stefan Willitsch}
\email{stefan.willitsch@unibas.ch}
\affiliation{Department of Chemistry, University of Basel, 4056 Basel, Switzerland}

%\phone{+41 61 207 38 30}

\date{\today}

%\singlespace

%\begin{tocentry}
%   \centering
%   \includegraphics{TOC_entry.pdf}
%\end{tocentry}

\begin{abstract}\noindent% 
The Diels-Alder cycloaddition, in which a diene reacts with a dienophile to form a cyclic compound, counts among the most important tools in organic synthesis. Achieving a precise understanding of its mechanistic details on the quantum level requires new experimental and theoretical methods. Here, we present an experimental approach that separates different diene conformers in a molecular beam as a prerequisite for the investigation of their individual cycloaddition reaction kinetics and dynamics under single-collision conditions in the gas phase. A low- and high-level quantum-chemistry-based screening of more than one hundred dienes identified 2,3-dibromobutadiene (DBB) as an optimal candidate for efficient separation of its \emph{gauche} and \emph{s-trans} conformers by electrostatic deflection. A preparation method for DBB was developed which enabled the generation of dense molecular beams of this compound. The theoretical predictions of the molecular properties of DBB were validated by the successful separation of the conformers in the molecular beam. A marked difference in photofragment ion yields of the two conformers upon femtosecond-laser pulse ionization was observed, pointing at a pronounced conformer-specific fragmentation dynamics of ionized DBB. Our work sets the stage for a rigorous examination of mechanistic models of cycloaddition reactions under controlled conditions in the gas phase.
\end{abstract}


\maketitle

%\tableofcontents
\section{Introduction}
\label{sec: introduction}

Besides polar and radical reactions, pericyclic processes are one of the three fundamental reaction types that form the basis of synthetic organic
chemistry. Given their great significance in organic synthesis, rigorously defined mechanistic
pathways are an important resource for reaction developers. The Diels-Alder
cycloaddition~\cite{diels28a}, in which a diene and a dienophile react to form a cyclic product, is
a practical and widely used pericyclic reaction in organic synthesis. While the broad strokes of its
mechanism are well understood, the detailed reaction manifold for any given substrate is often
extensively discussed~\cite{houk95a, serafimov08a, domingo09a, domingo14a, souza16a, rivero19a}. Given its
mechanistic subtleties and its importance, the Diels-Alder reaction has served as a test-bed for
establishing new types of mechanistic analysis \cite{singleton95a}. In case of the ``canonical''
concerted pathway, which involves a cyclic transition state and is widely discussed in
the literature~\cite{houk95a, serafimov08a, domingo09a, domingo14a, souza16a, rivero19a}, only the \emph{s-cis}
conformer of the diene reacts to form the cyclic product -- while in a stepwise mechanism
also the \emph{s-trans} conformer can contribute to the reaction. Stepwise pathways become
particularly important for ionic variants of the reaction, i.e., polar cycloadditions, for which
traditional concepts for rationalizing the mechanism such as the conservation of orbital symmetry
break down~\cite{donoghue06a, domingo09a, rivero17a, rivero19a}. Thus, an experimental investigation of the
mechanistic impact of individual rotamers is clearly warranted. This, however, requires a way to
probe the reactivities of the individual conformers of a specific diene, a difficult task under
standard liquid-phase reaction conditions.

In recent years, molecular beams have become an important tool for the investigation of gas-phase
chemical reaction dynamics under highly controlled conditions\cite{chang15a,willitsch17a}. In
particular, the use of inhomogeneous electric fields has enabled the electrostatic deflection and thus spatial separation of different molecular
conformers and isomers according to their different electric dipole
moments\cite{Filsinger:PRL100:133003, filsinger09a, kierspel14a, chang13a, kilaj18a}. The
combination of such a ``controlled'' molecular beam with a stationary reaction target of
sympathetically cooled molecular ions in an ion trap forms a powerful tool for studies of the
kinetics and dynamics of ion-molecule reactions~\cite{chang13a, roesch14a, willitsch17a, kilaj18a}.
Recently, this approach has enabled the measurement of individual chemical reactivities of the
\emph{cis} and \emph{trans} conformers of 3-aminophenol with trapped Ca$^{+}$ ions \cite{chang13a,
   roesch14a} and of the two nuclear-spin isomers of water~\cite{horke14a} toward trapped
diazenylium ions (N$_2$H$^+$) in a proton-transfer reaction \cite{kilaj18a}. Consequently, molecular
beams in conjunction with ion traps offer a direct and precise way to measure conformer-specific
rate constants and thus to investigate the reaction mechanism of polar cycloadditions. The key
challenge is the identification of suitable model systems amenable to a characterisation under these
specific experimental conditions. In the present context, this means that (i) both reactants, the
diene and the dienophile, need to be volatile enough to enable their preparation in the gas phase,
(ii) the energy difference between the \emph{s-cis} and \emph{s-trans} conformers of the diene needs
to be small enough so that both can be populated in the cold environment of a supersonic molecular
beam, and (iii) the difference of their permanent dipole moments in the molecular frame needs to be
large enough to enable their efficient electrostatic separation \cite{chang13a, chang15a}.

The selection of an optimal model system is, therefore, a multi-dimensional optimization problem.
Traditionally, the choice would be guided by chemical and physical insight. Nowadays, numerical
simulations of reactions can provide meaningful atomistic insights to support experimental
efforts~\cite{sperger16a}. In the context of designing experiments, virtual screening has proven to be
a powerful approach for suggesting compounds matching the physical and chemical properties of
interest. Computational screening has already been successfully applied in protein, materials and
catalytic design~\cite{notz03a, er15a, doney16a}. Here, we apply this approach to identify optimal
dienes suitable for controlled gas-phase polar cycloaddition reactions.

%\paragraph{Outline}
This article combines methods of theoretical, organic, and physical chemistry to lay the foundations
for a subsequent experimental characterization of conformational effects in polar cycloaddition reactions.
Quantum-chemical calculations were performed to screen the reactant space for a diene with physical
properties optimized toward its use in conformer-selected reaction-dynamics experiments in the gas
phase. A synthesis for the theoretically identified optimal diene, 2,3-dibromobutadiene (DBB), was
then developed. Finally, the physical properties of the compound were validated in a molecular-beam
experiment separating the two conformers by electrostatic deflection.

\section{Theoretical and experimental methods}
\label{sec:methods}

\subsection{Theoretical screening}
\label{sec:screening}

We applied the concept of computational screening towards the problem of exploring chemical space
from first principles~\cite{lilienfeld13a} in order to find a polar diene for conformer-selective
Diels-Alder cycloadditions. The chemical role of polar dienes in Diels-Alder reactions was already
explored computationally in various preceding studies, see, e.g., references~\cite{dewar86a,
   rivero17a}. Efficient electrostatic separation necessitates a certain difference in electric
dipole moments $\Delta\mu$ between the \emph{s-cis} and \emph{s-trans} conformers of the
diene~\cite{Filsinger:PRA82:052513, chang15a}. Moreover, a small energy difference $\Delta{E}$
between the ground states of the conformers is required to ensure a significant thermal population
of both species in the molecular beam. For a successful experiment, a suitable diene has to be
identified which satisfies both of these conditions.

High-throughput based virtual design of novel compounds typically starts from an initial scaffold,
which can easily be modified at multiple sites through functionalization by substituting atoms or
functional groups~\cite{lilienfeld14a}. Theoretical screening then yields the best mutated
combinations selected according to their proximity to the desirable physical or chemical
target-property values.

In the present work, we computationally searched the chemical space of butadiene derivatives for
which the \emph{s-cis} and \emph{s-trans} isomers exhibit maximal and minimal differences in dipole
moments and energies, respectively. Substituting CH$_2$ in positions 1 and 4 by NH, or O, and
substituting the hydrogen attached to carbon in position 2 and 3 by halogens (F, Cl, Br, I), a
preliminary density functional theory (DFT) based scan of 144 candidates, not accounting for
symmetrically redundant species, resulted in the identification of di-halogen substituted butadiene
as a promising series of candidates for experiments. Due to the chemical reactivity of
iodine-substituted compounds, potentially hampering subsequent synthetic efforts, we have only
included the difluoro, dichloro, and dibromo 2,3-substituted butadienes for further in-depth
theoretical analysis.

Torsional energy profiles were subsequently calculated for all three species using DFT with the
double-hybrid functional DSD-PBEP86-D3BJ\cite{kozuch11a} and a large basis set
(def2-QZVPP)\cite{weigend05a} which was previously shown to give good performance for the
prediction of torsional potential energy surfaces of similar molecules.\cite{tahchieva18a} For the
torsional profiles, the geometry optimizations were restricted by keeping the torsional angle
$\Theta = \Theta_{\textrm{H}_3\textrm{C}-\textrm{CC}-\textit{Y}}$ constant, imposing achirality. The
entire range of 0$^{\circ} < \Theta < 180^{\circ}$ was scanned in steps of $\Delta \Theta$ =
20$^{\circ}$. Note that due to the applied constraints for the
torsional angles throughout the geometry optimization, the torsional profile is symmetric
[$E(360^{\circ} - \Theta) = E(\Theta$)]. Calculations were carried out with the Gaussian09 program
package~\cite{g09}.

\subsection{Stark-energy and trajectory simulations}
To theoretically assess the behaviour of DBB in an electrostatic deflection experiment, Stark energies and effective dipole moments of individual rotational states of DBB were calculated
using the CMIstark software package\cite{chang14a}. The calculated rotational constants and dipole
moments as listed in \autoref{tab:molecular_constants} were used as input parameters. The Stark
energies served as input parameters for simulating state-specific deflection profiles of the
molecular beam by the electrostatic deflector with a home-made software package based on CMIfly
\cite{chang15a}.

Trajectory simulations were carried out for \emph{gauche}-DBB with $10^5$ molecules for each
rotational state up to a maximum rotational quantum number of $J_\mathrm{max}=20$. For the unpolar
\emph{s-trans}-DBB, only a single quantum state, $J=0$, needed to be simulated with a total
number of $10^6$ trajectories. In all cases, initial positions were uniformly sampled across the
cross section of the orifice of the gas nozzle generating the molecular beam. The initial velocities
of the molecules were sampled from a normal distribution. The velocity distribution was matched to
the experimentally determined mean longitudinal velocity of 843~m/s with a longitudinal velocity
spread of 10~\%. A transverse velocity spread of 4~m/s was chosen to match the divergence of the
beam to the acceptance angle of the skimmers in the assembly. According to the theoretical energy
difference between the ground states of \emph{gauche}- and \emph{s-trans}-DBB in
Table~\ref{tab:molecular_constants}, the ratio of their thermal populations at room temperature is
$p_\mathrm{gauche}/p_\mathrm{trans}=0.30$, taking into account the two-fold degeneracy of the
\emph{gauche} structure, see \autoref{fig:torsional_profile_br}. This ratio was used to scale the
simulated deflection profiles of the two species.

In order to calculate thermally averaged deflection profiles $n_{\sigma, T}(y)$ for each conformer
($\sigma \in \{\text{\emph{gauche}},\text{\emph{s-trans}}\}$) at a specific rotational temperature
$T$, we followed a similar procedure as before \cite{Filsinger:JCP131:064309, kilaj18a}. For each
rotational quantum state $|J_{K_a K_c} M\rangle$, histograms of the arrival positions
$n_{\sigma JK_aK_cM}(y)$ normalized by the initial sample size were extracted from the simulated
trajectories. Here, $J$ is the quantum number for overall angular momentum neglecting nuclear spin,
i.e., for the overall rotation, $K_a$ and $K_c$ are pseudo-quantum numbers for the projection of the
angular momentum onto the molecular axes, and $M$ is the quantum number for the projection of the
rotational angular momentum onto the external-field axis. Thermal averaging was performed using the
relation
\begin{eqnarray}
   n_{\sigma, T}(y) &=& \frac{p_\sigma}{N_\sigma} \sum_{J=0}^{J_\mathrm{max}}\sum_{K_a, K_c} \sum_{M=0}^{J} \times\\
   & & \times g_{M} e^{-E_{JK_aK_c}/k_B T} \; n_{\sigma JK_aK_cM}(y) ,
\end{eqnarray}
with the partition function
\begin{equation}
   N_\sigma = \sum_{J=0}^{J_\mathrm{max}}\sum_{K_a, K_c} \sum_{M=0}^{J} g_{M} e^{-E_{JK_aK_c}/k_B T}.
\end{equation}
Here, $k_B$ denotes the Boltzmann constant, $p_\sigma$ are the populations of the conformers at room
temperature and $E_{JK_aK_c}$ are the field-free rotational energies. The degeneracy factor $g_M$
takes values $g_M = 1$ for $M=0$ and $g_M = 2$ for $M>0$. The total thermal deflection profile was
calculated from the sum of the deflection profiles of the \emph{gauche}- and
\emph{s-trans}-conformers,
\begin{equation}
   n_\mathrm{tot, T}(y) = n_{\mathit{gauche}, T}(y) + n_{\mathit{s-trans}, T}(y).
\end{equation}

\subsection{Synthesis of DBB}
Since DBB is an unstable compound which is not commercially available, we needed to devise a
synthesis that delivered the material in sufficiently high purity and quantity for molecular-beam
experiments. Prior to this work, \citet{stewart62a} reported a synthesis of DBB from
1,4-dihalo-2-butyne and a cuprous halide which formed activated halide ions present in solution. DBB
was obtained through continuous distillation during the reaction. After extensive screening of
potential conditions, we found that the elimination reaction of 1,2,3,4-tetrabromobutane (TBB) with
the sterically hindered base 1,8-Diazabicyclo[5.4.0]undec-7-ene (DBU) primarily delivered the
elimination product DBB. The most effective conditions involved adding DBU to a solution of TBB in
diethyl ether under a constant stream of nitrogen, with NaI as an additive to accelerate the
substitution. Under these conditions, near complete conversion to DBB was observed after 1 hour of
reaction time. The purified sample was directly used in the molecular beam apparatus. Further
information on the synthesis can be found in the electronic supplementary information (ESI).

%\subsection{Experimental setup}

\subsection{Experimental setup for conformer separation}
\begin{figure*}[!t]
   \includegraphics[width=\linewidth]{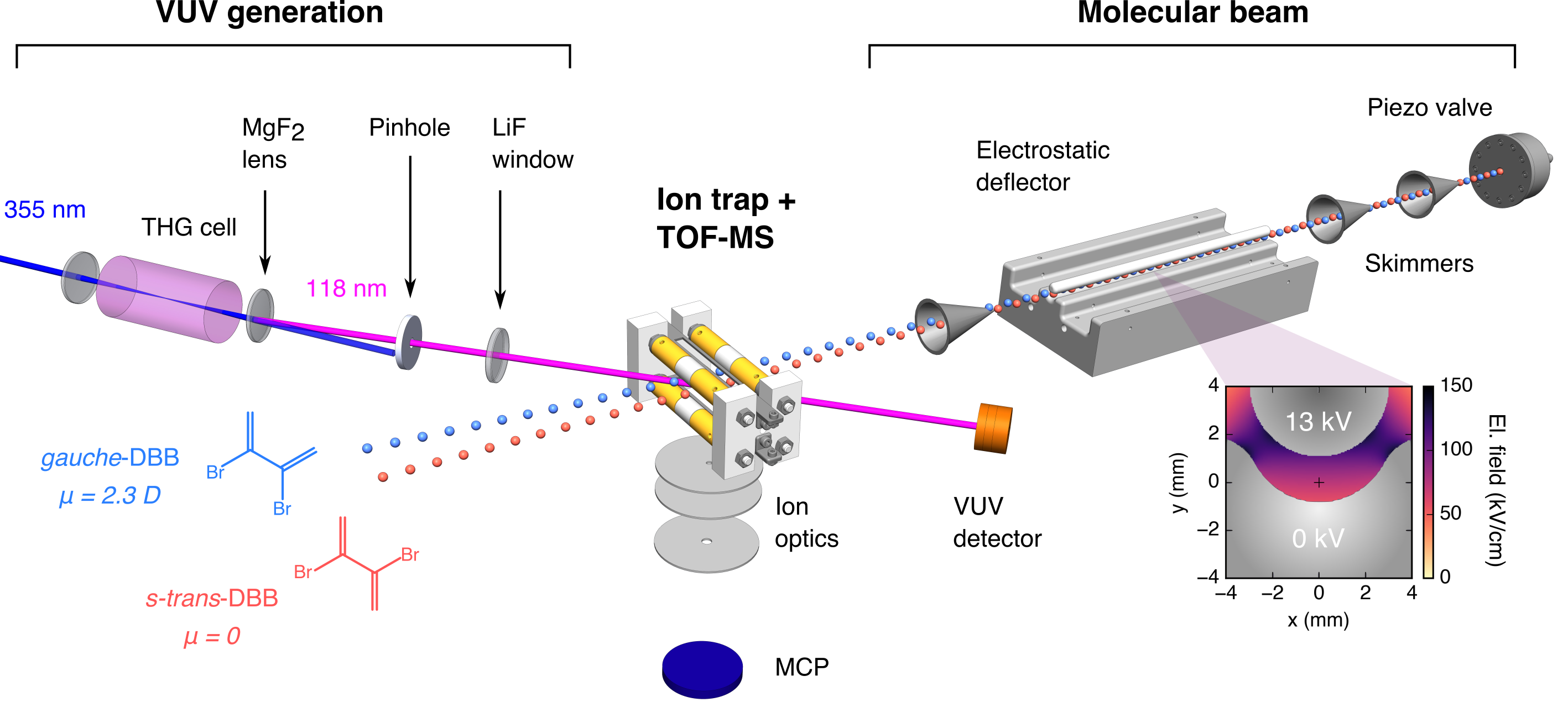}
   \caption{\textbf{Overview of the experimental setup.}}
   \label{fig:setup}
\end{figure*}

A schematic of the experiment is depicted in \autoref{fig:setup}. Details of the experimental setup have also been described in our earlier work \cite{chang13a, roesch14a, roesch16a, kilaj18a}.

A supersonic jet of DBB seeded in neon was generated using a pulsed gas valve and passed through two
skimmers before entering the electrostatic deflector. The resulting molecular beam contained a
mixture of the \emph{gauche} and \emph{s-trans} conformers of DBB. The inset of \autoref{fig:setup}
depicts the inhomogeneous electric field in the deflector with a cross marking the nominal molecular
beam axis. Here, the two conformers were angularly dispersed and thus spatially separated according
to their different dipole moments~\cite{chang15a}. Behind the deflector, the molecular beam was
directed at a linear-quadrupole ion-trap (LQT) coupled to a time-of-flight mass spectrometer
(TOF-MS). The entire molecular beam setup can be tilted vertically with respect to the TOF-MS, which
allows probing different regions of the dispersed molecular beam. The tilting angle thus defines a
deflection coordinate $y$. When entering the TOF-MS, the DBB molecules were ionized by either pulsed
vacuum-ultraviolet (VUV) radiation or femtosecond (fs) laser pulses and accelerated onto a
microchannel-plate detector (MCP) using high-voltage electrodes.

\subsubsection{Molecular beam:} % molecular beam details
The molecular beam was generated from DBB vapor at room temperature seeded in neon carrier gas at
5~bar. The gas mixture was pulsed through a cantilever piezo valve (MassSpecpecD ACPV2, $150~\mu$m
nozzle diameter) at a repetition rate of 10~Hz and a gas pulse duration of $250~\mu$s at the LQT. The velocity in the direction of propagation of the resulting molecular beam was measured in the same way as described in \citet{kilaj18a} and yielded a value of
843(58)~m/s. To determine the beam density of DBB in the molecular beam, we first calibrated the
absolute sensitivity of the TOF-MS by loading Coulomb crystals of defined ion number in our ion trap
\cite{schmid17a}. Then, following refs.~\citealp{hankin01a, wiese19a}, we measured the total ion
yield of DBB from the molecular beam using ionization with 150~fs, 775~nm laser pulses. Observing
the logarithmic increase of ion yield with laser intensity~\cite{wiese19a} allowed us to
estimate a DBB density of $7.8(5)\times10^7~\mathrm{cm}^{-3}$. For deflection experiments, a voltage
of 13 kV was applied between the two 15.4~cm long deflector electrodes, held at a distance of
1.4~mm, to generate the required vertical electric field gradient~\cite{Filsinger:JCP131:064309,
   chang15a, roesch14a, kilaj18a}. By comparison of the measured deflection profiles with
Monte-Carlo simulations~\cite{Filsinger:JCP131:064309, chang13a, kilaj18a}, a rotational temperature of 1.0~K could be
estimated.

\subsubsection{Ion trap and TOF-MS:} % Trap and TOF-MS
The LQT is connected to a TOF-MS orthogonal to the molecular-beam propagation axis for quantitative
mass analysis\cite{roesch16a}. To enable a better ion selection in the TOF-MS, a voltage of 500 V
was applied to the four end caps of the trap. For extraction of the ionized species, a permanent
voltage of 4.0~kV was applied to the repeller electrode. A microchannel plate detector (MCP,
Photonis USA) operating at a typical voltage of 2.3 kV was placed at the end of the flight tube.

\subsubsection{Femtosecond and VUV ionization:} % Fs laser ionization and VUV generation
Non-resonant multi-photon ionization of the DBB molecules was performed with pulses from a
Ti:Sapphire femtosecond laser (CPA 2110, Clark-MXR, Inc.) at a wavelength of 775~nm and pulse
duration of 150~fs focused to a diameter of $\sim30~\mu$m in the sample. In addition, a
vacuum-ultraviolet (VUV) light source was used for soft ionization of the DBB molecules. Similar to
other work~\citep{shi02a, vanbramer92a, steenvorden91a}, pulses of 118~nm light were generated using
third-harmonic generation (THG) by focusing the third harmonic output beam of a Nd:YAG laser
(Quantel Brilliant, 355~nm, 5~ns) into a gas cell containing a phase-matched gas mixture of xenon and
argon (ratio 1:10, total pressure 100~mbar). The pump laser was operated at a repetition rate of
10~Hz and the pulse energy was set to 25~mJ such that a UV to VUV conversion efficiency of
approximately $10^{-5}$ was achieved. The VUV beam was re-focused (spot size $\sim100~\mu$m) into the trap chamber at the center of the ion trap with a single MgF2 lens (Thorlabs,
f = 200 mm) at a distance of 120 mm from the UV focus (spot size $\sim15~\mu$m). Owing to its stronger index of refraction in
the VUV, the MgF$_2$ lens also served as an optical element to separate the pump-laser beam from the
VUV beam. A LiF window was used to seal off the ultra-high-vacuum chamber housing the ion trap from
the VUV generation chamber. In order to block the 355 nm pump-laser beam and prevent it from
entering the interaction region or damaging the UV sensitive LiF window, a MACOR-protected pinhole
was installed in front of the LiF window. The VUV detector was fabricated from two copper electrodes
with a typical bias voltage of about 1~kV and the VUV-induced photocurrent was measured through the
resulting voltage across a $50~\Omega$ resistor.

\section{Results and discussion}
\label{sec: results}

\subsection{Torsional profiles of the 2,3-dihalobutadienes}
\label{sec:res_screening}

\begin{figure}[!t]
    \centering
    \includegraphics[width=\linewidth]{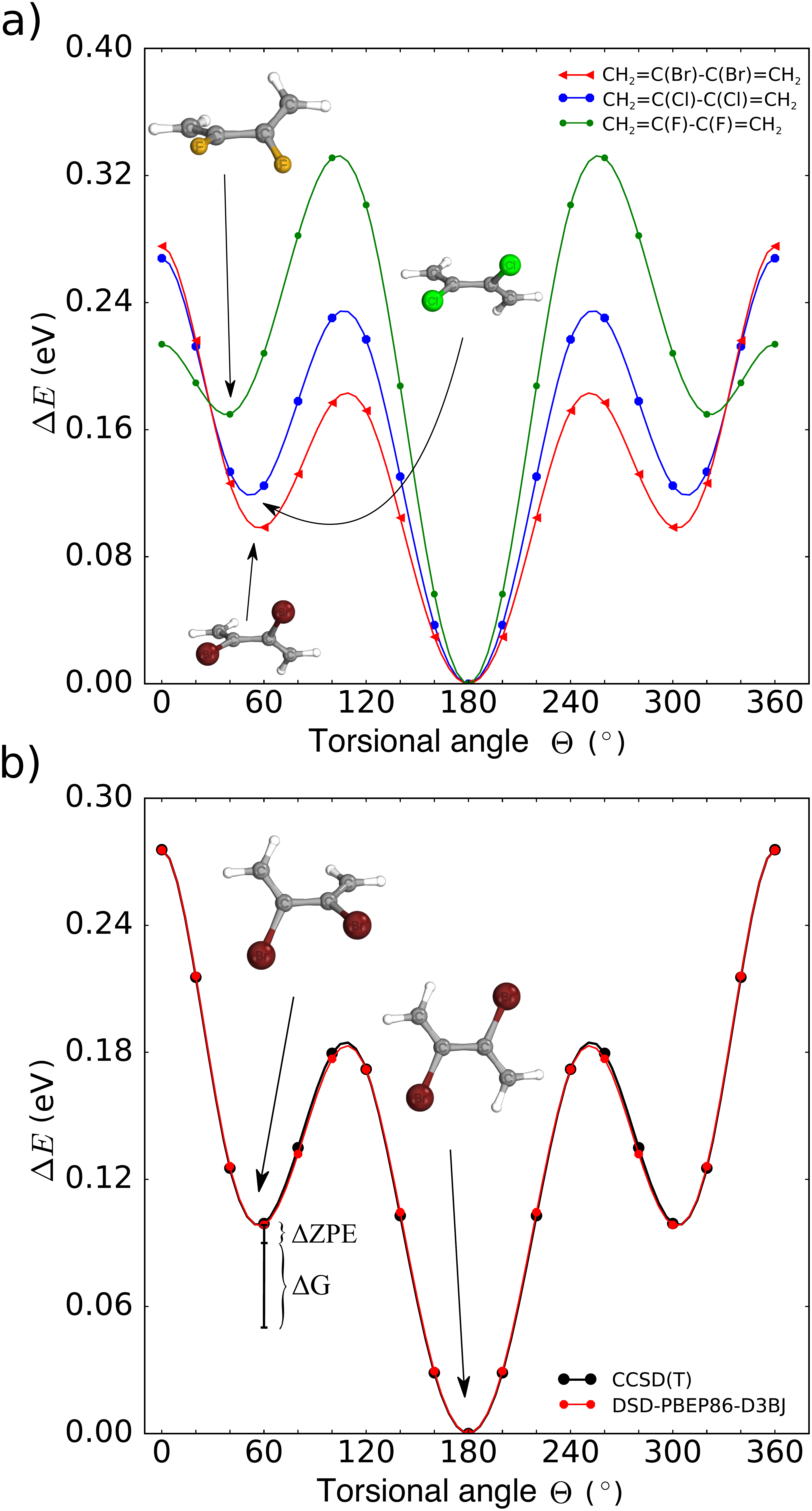}
    \caption{a) Cuts through the potential energy surface of 2,3-dibromo-1,3-butadiene (CH$_2$=C(Br)-C(Br)=CH$_2$),
       2,3-dichloro-1,3-butadiene (CH$_2$=C(Cl)-C(Cl)=CH$_2$), and 2,3-difluoro-1,3-butadiene
       (CH$_2$=C(F)-C(F)=CH$_2$) along the torsional coordinate $\Theta$ calculated using the DSD-PBEP86-D3BJ functional. Due to the symmetry of the
       molecules, the torsional profiles are symmetric with respect to mirroring at
       $\Theta=180\,^\circ$. The figure shows relative energies $\Delta E$ referenced to the energies of the \emph{s-trans} structures at $\Theta=180^\circ$. b) Potential energy of 2,3-dibromo-1,3-butadiene as a function of the torsional angle $\Theta$
       calculated at CCSD(T)/VTZ-F12 and DSD-PBEP86-D3BJ/def2-QZVPP levels of theory. The lowering
       of the energy at the local minima due to the zero point vibrational energy
       ($\Delta\text{ZPE}$) and the Gibbs free energy ($\Delta G$) is illustrated for the
       $gauche$ structure. Due to its symmetry, the molecule exhibits two equivalent \emph{gauche}
       structures.}
    \label{fig:torsional_profile_br}
\end{figure}

A graphical representation of the torsional profiles of the 2,3-difluoro-, dichloro-, and dibromobutadienes is shown in \autoref{fig:torsional_profile_br} (a). The global minimum was found to be the \emph{s-trans} structure (at $\Theta = 180^\circ$) in all cases. Local minima were found to be near \emph{gauche} (rather than \emph{s-cis}) structures at torsional angles varying from $\Theta=50^\circ$ to 60$^\circ$ depending
on the specific molecule.

Subsequently, differences in potential energy and absolute dipole moment between the local and
global minima were calculated including harmonic and anharmonic thermal corrections of
zero-point vibrational energy and Gibbs free energy. The DFT results for the torsional potential and
the dipole moments were in very good agreement with
CCSD(T)-F12/cc-pVTZ-F12,\cite{werner11a,peterson08a} and CCSD/cc-pVTZ-F12\cite{purvis82a}
(\autoref{tab:therm_corr}) calculations, respectively. Furthermore, relaxation at the
CCSD/cc-pVTZ-F12 level resulted in geometries identical to those found by DSD-PBEP86-D3BJ/def2-QZVPP
with a root-mean-square deviation (RMSD) of 1.9~pm between the final geometries. These results
confirm the reliability of our DFT predictions.

While exhibiting still non-negligible conformational energy differences, see
\autoref{tab:therm_corr}, the large dipole moment differences among the 2,3-di[halogen]but-1,3-diene
conformers appeared promising, motivating its selection for subsequent experimental investigations.
For instance, the dipole-moment difference for 2,3-dibromo-1,3-butadiene was computed at the CCSD/cc-pVTZ-F12
level of theory (neglecting all relativistic effects) to be $\Delta\mu=2.11$~D.
\begin{table*}[!t]
\centering
\begin{adjustbox}{width = \linewidth}%scale=0.60
\begin{tabular}{c||c|c|c}
 %\hline
   Method & CH$_2$=C(Br)-C(Br)=CH$_2$  & CH$_2$=C(Cl)-C(Cl)=CH$_2$ &  CH$_2$=C(F)-C(F)=CH$_2$ \\ \hline \hline
       & \multicolumn{3}{c}{$\Delta E$ (eV)} \\ \hline
    CCSD(T)/cc-pVTZ-F12  & 0.097  &  0.117  &  0.151  \\
   DSD-PBEP86-D3BJ/def2-QZVPP  & 0.097 & 0.114 & 0.155   \\
   DSD-PBEP86-D3BJ/def2-QZVPP + harm. therm. corr. & 0.050 & 0.069 & 0.142  \\
   DSD-PBEP86-D3BJ/def2-QZVPP + anharm. therm. corr. & 0.049 & 0.068 & 0.139 \\ \hline%\hline
     & \multicolumn{3}{c}{$\Delta\mu$ (Debye)} \\ \hline
    CCSD/cc-pVTZ-F12 & 2.1072  & 2.2831  & 2.5828  \\
    DSD-PBEP86-D3BJ/def2-QZVPP & 2.2963 & 2.3837 & 2.5938 \\
\end{tabular}
\end{adjustbox}
\caption{Differences in potential energy $\Delta E$ and dipole moment $\Delta\mu$ between the local
   and global torsional minimas of DBB including thermal corrections at T=298.15K (zero-point and
   Gibbs free energy) for selected 2,3-dihalogen-substituted butadienes. }
\label{tab:therm_corr}
\end{table*}

As the main result of the theoretical screening, 2,3-dibromobuta-1,3-diene (DBB) was identified as
an optimal diene for the envisaged experiments that possesses both a sufficiently small energy gap
between the \emph{gauche} and \emph{s-trans} ground states as well as a large enough difference in
the electric dipole moment of the two species (\autoref{fig:torsional_profile_br} (b) and \autoref{tab:molecular_constants}) . For \emph{gauche}-DBB a dipole moment of $\mu=2.29$~D
was calculated at the DSD-PBEP86-D3BJ/def2-QZVPP level of theory, while the \emph{s-trans} isomer is
apolar on grounds of its inversion symmetry. \autoref{tab:molecular_constants} summarizes the
calculated energy difference as well as the absolute values of the dipole moments and the rotational
constants for both conformers.
\begin{table*}[!t]
\centering
\begin{adjustbox}{scale = 1}
\begin{tabular}{l || c | c | c c c}
	&  Energy 	& Dipole moment  	& \multicolumn{3}{c}{Rotational constants (GHz)}\\
	&  $\Delta E = E_\mathrm{cis} - E_\mathrm{trans}$ (eV)		& $\mu$ (D)		&${A_e}$ 		& ${B_e}$ 	& ${C_e}$\\ \hline\hline
%\emph{s-cis} DBB 	& \multirow{2}{1 cm}{0.026} 		& 2.183 			& 0.70058 	& 0.86858 	& 2.31501\\
\emph{gauche}-DBB 	& \multirow{2}{1 cm}{0.049} 		& 2.29 			& 2.3526 	& 0.8793 	& 0.7097\\
\emph{s-trans}-DBB 	& \multirow{2}{1 cm}{} 		&  		0.00	& 4.6077 & 0.5997 &  0.5306 \\
%\emph{s-trans} DBB 	&  			& 0 				& 0.52311 	& 0.59121 	& 4.54110
\end{tabular}
\end{adjustbox}
\caption{Differences in energy, dipole moments and rotational constants of \emph{gauche}- and
   \emph{s-trans}-2,3-dibromobuta-1,3-diene (DBB) calculated at the DSD-PBEP86-D3BJ/def2-QZVPP level
   of theory including anharmonic thermal corrections.}
\label{tab:molecular_constants}

\end{table*}

\subsection{Simulations of the electrostatic deflection of DBB}
\begin{figure*}[!t] % simulated stark energies and deflection profiles
   \centering
   \includegraphics[width=0.8\linewidth]{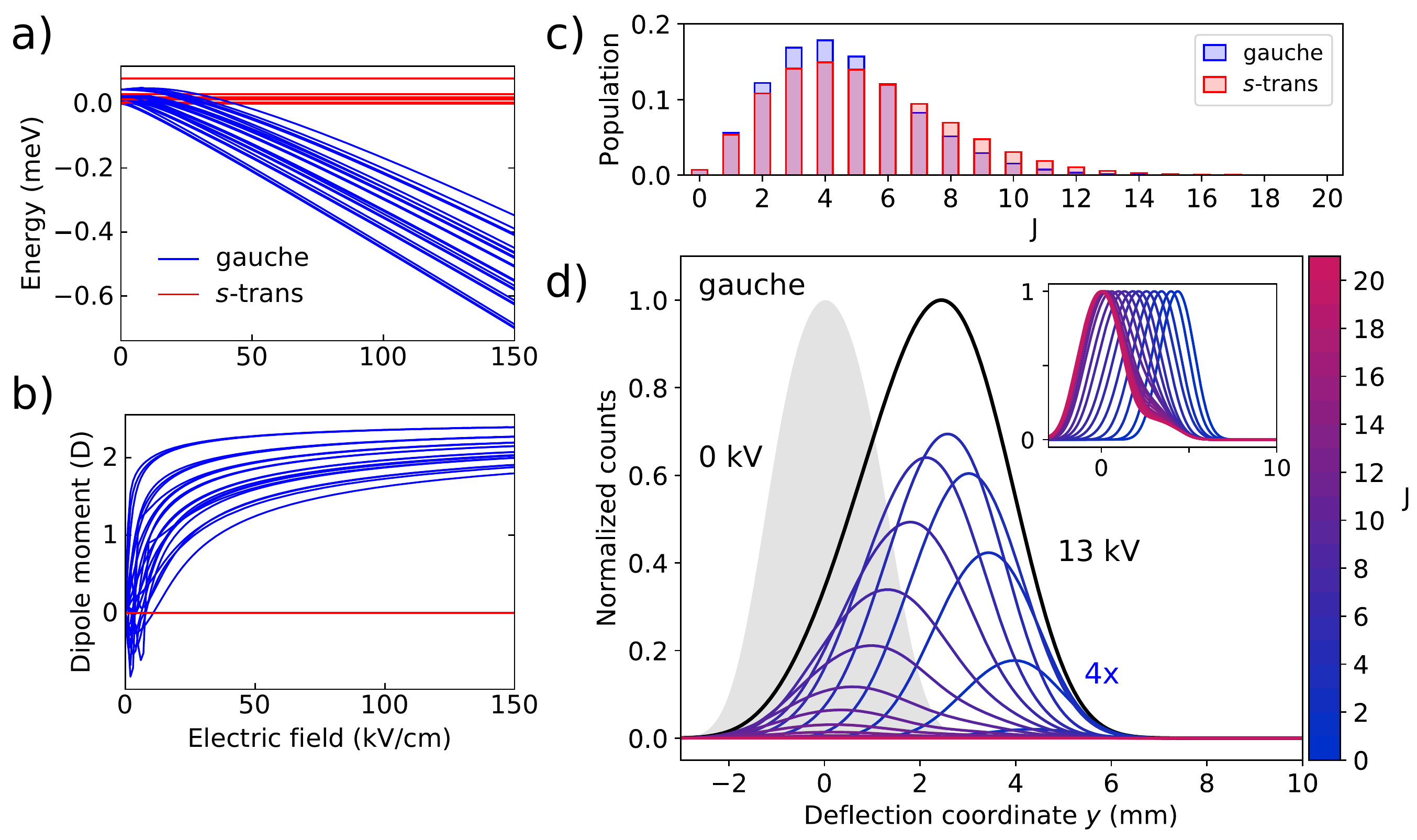}
   \caption{\textbf{Simulations.} Calculated Stark energies a) and effective dipole moments b)
      \emph{vs.} electric field strength for individual rotational states with $J=0,1,2$ of the \emph{gauche} and \emph{s-trans}
      conformers of DBB. c) Rotational state populations, summed over all levels with the same
      angular momentum quantum number $J$, for \emph{gauche} and \emph{s-trans} DBB at a rotational
      temperature of 1.0~K. d) Simulated deflection profile of the \emph{gauche} conformer (black
      line) with its different rotational-state contributions in color. The contributions of the different $J$ states are color-coded according the color scale indicated. The undeflected beam profile is shown by the gray area.}
   \label{fig:stark_simulation}
\end{figure*}
Based on the molecular properties obtained from the computations, we predicted trajectories of
\emph{gauche}- and \emph{s-trans}-DBB molecules through the electrostatic deflector. In
\autoref{fig:stark_simulation}, calculated Stark energies (a) and effective dipole moments (b) for
rotational states with angular momentum quantum numbers up to $J=20$ of the \emph{gauche} and
\emph{s-trans} conformers of DBB are shown as a function of electric field strength. In the applied
electric fields, all rotational states of the \emph{gauche} conformer are strong-field seeking with
negative Stark shifts, whereas the \emph{s-trans} conformer does not exhibit a DC Stark effect
because of its vanishing dipole moment in the molecular frame. In \autoref{fig:stark_simulation}~c),
rotational state populations for \emph{gauche}- and \emph{s-trans}-DBB at a rotational temperature of
1.0~K are shown. At this temperature, rotational states up to $J=14$ are significantly populated and
can be expected to contribute to the beam-deflection profiles for both conformers.

The density profiles of the molecular beam along the deflection coordinate at the position of
intersection with the probe laser (deflection profiles) are plotted in
\autoref{fig:stark_simulation}~d) for the \emph{gauche} conformer. The color-coded curves show the
contributions from the individual rotational states with angular momentum up to $J=20$, while the
thick black line corresponds to the total thermally averaged deflection profile at a rotational
temperature of 1.0~K. For clarity, the contributions of the individual rotational states have been
multiplied by a factor of 4 in the figure. The inset contains the same curves with heights
normalised to 1 to allow for a better comparison. The grey area in the main plot is a simulation of
the undeflected beam profile, at a deflector voltage of 0~kV, which also corresponds to the profile
of the unpolar \emph{s-trans} conformer with the deflector voltages turned on. The rotational states
of the \emph{gauche} conformer with largest deflection are the low-angular-momentum states (small
$J$). Consequently, significant spatial separation of the \emph{gauche} and \emph{s-trans}
conformers can only be achieved experimentally for samples with a sufficiently low rotational
temperature~\cite{Trippel:RSI89:096110}.

\subsection{Experimental deflection profiles}

\begin{figure}[!t]
   \centering
   \includegraphics[width=\linewidth]{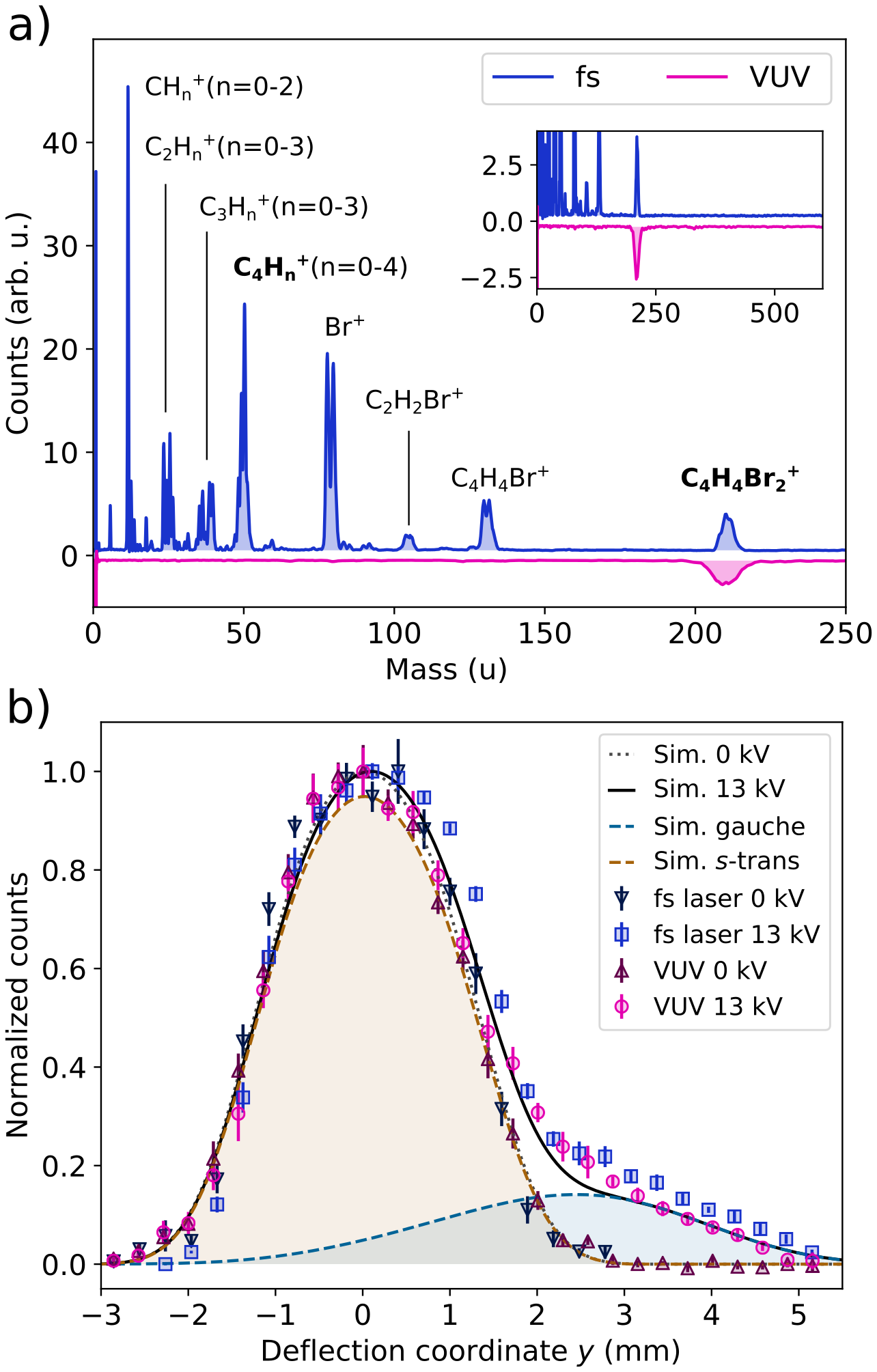}%deflection
   \caption{\textbf{Deflection profiles.} a) Representative TOF-MS trace obtained using
      fs-laser-pulse ionization (top) and VUV ionization (bottom) of DBB. b) Molecular-beam profiles
      measured using both ionization methods at deflector voltages of 0 kV and 13 kV together with
      corresponding simulations. For fs-laser-pulse ionization, the profiles for the fragment
      C$_4$H$_n^{+}$ are shown. Error bars represent standard errors of at least five independent
      measurements. }
   \label{fig:deflection}
\end{figure}

In order to measure the spatial profiles of the DBB molecules beam emanating from the electrostatic deflector,
the molecules were ionized by laser pulses and ejected into the TOF-MS. The choice of the ionization
method turned out to be crucial. In \autoref{fig:deflection}~a), typical TOF-MS traces obtained using
fs-laser-pulse ionization (top) and VUV ionization (bottom) are shown. While fs-laser
ionization yielded a large quantity of fragmentation products of the parent DBB molecule, VUV
ionization produced a clean mass spectrum with a single peak originating from DBB at 212~u. The
inset shows an extended mass range around the DBB peak, illustrating that other species or clusters
with larger mass cannot be observed under the present experimental conditions.

\begin{figure}[!t]
\centering
   \includegraphics[width=\linewidth]{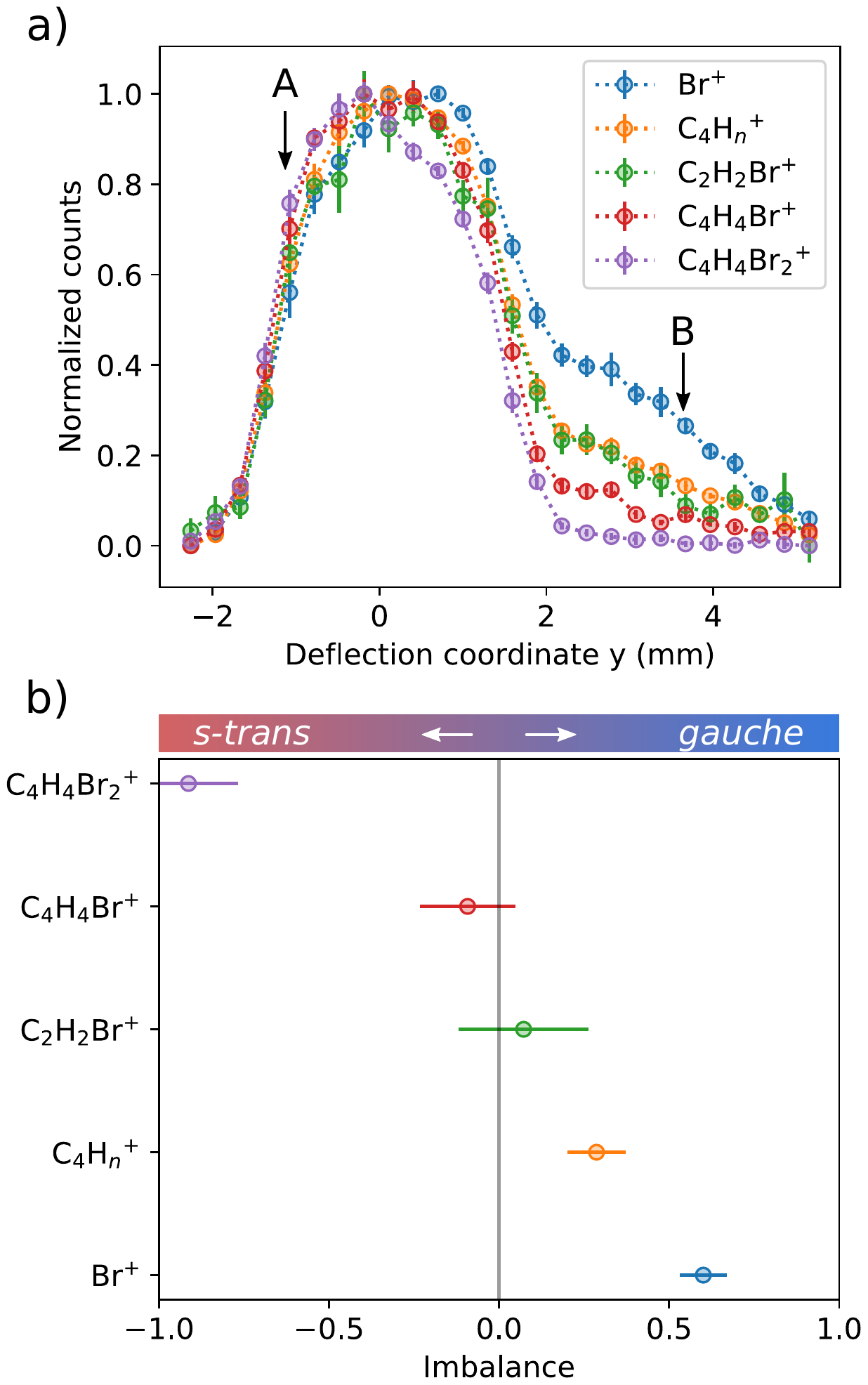} % deflection
   \caption{\textbf{Deflection curves of a molecular beam containing DBB probed at different ionic
         fragment masses produced by fs-laser-pulse ionization.} a) Ion counts of four molecular
      fragments and the mass of the parent DBB molecule vs. deflection coordinate. b) Measured
      imbalance of fs-laser ionization products between the two points A and B in a). See text
      for details. Error bars represent standard errors of six independent measurements. }
   \label{fig:imbalance}
\end{figure}

Analysis of the different fragment-ion signals obtained from fs-laser ionization,
\autoref{fig:imbalance}~a), revealed that most of the fragments show distinct deflection profiles.
Intriguingly, the mass signal corresponding to the parent molecule does not seem to exhibit
deflection. This signal could in principle be generated by the break up of larger DBB-containing
clusters which may exhibit only very small dipole moments, similar to the situation observed in the
deflection of H$_2$O \cite{kilaj18a}. However, DBB cluster ions are not observed in the TOF spectra,
\autoref{fig:deflection}~a), and hence we can essentially rule out that the lack of deflection
observed for the DBB$^{+}$ mass peak measured by fs ionization is due to breakup of molecular
aggregates. The different deflection profiles recorded for ion signals of the individual fragments
are caused by the fs-laser-induced breakup of the parent DBB molecule. It is possible that the
electric field of the relatively long laser pulses drives different multiple-ionization dynamics for
the polar \emph{gauche} conformation than for to the apolar \emph{s-trans} conformation, thus
leading to distinct conformer-specific fragmentation patterns~\cite{zigo17a}; these are further
discussed on the next page. 

The complexity of the observed fragmentation dynamics prevented us from unambiguously determining
the deflection profile of the DBB parent molecule using fs-laser ionization. Therefore, we
implemented soft VUV ionization, which is capable of ionizing DBB without fragmentation as apparent
from \autoref{fig:deflection}~a) and the corresponding molecular-beam profiles in
\autoref{fig:deflection}~b). While the data points measured with VUV (purple triangles and circles,
respectively) probe DBB directly, the data shown for fs-laser-pulse ionization corresponds to the
accumulated signal for the fragments $\mathrm{C_4H_n^{+}}~(n=0\ldots4)$ (blue triangles and squares,
respectively) produced under these conditions. The experimental data points for VUV ionization agree
very well with the simulated thermally-averaged beam profiles, which are shown as grey dotted line
(0 kV) and black solid line (13 kV). Corresponding individual contributions from the \emph{gauche}
and \emph{s-trans} conformers are depicted as the blue and orange shaded areas, respectively. The
deflection profile at 13~kV shows a tail towards higher deflection coordinates where simulations
indicate the presence of pure \emph{gauche}-DBB. The overall very good agreement between the
measured and simulated deflection profiles allows us to confirm the successful separation of the DBB
conformers and validates the accuracy of the theoretical
calculations.

Further evidence for the separation of the \emph{gauche} and \emph{s-trans} conformers can be found
in the measured fragmentation products due to fs-laser-pulse ionization of the molecular beam.
\autoref{fig:imbalance}~a) shows normalized profiles of four representative fragment families
Br$^+$, C$_4$H$_{4}^{+}$, C$_2$H$_n$Br$^+$, C$_4$H$_4$Br$^+$ and the parent molecule DBB$^+$ as a
function of the deflection coordinate. Clearly, the tail of the profile towards large deflection
coordinates, where one expects the contribution from \emph{gauche}-DBB, varies strongly among the
different fragments, with Br$^+$ showing the largest and DBB$^+$ almost zero amplitude. In the
region around $y \approx -1$~mm, this behavior is inverted. At this location, our trajectory
simulations predict a predominance of \emph{s-trans}-DBB. In order to quantify the imbalance of the
observed fragment yields for the \emph{gauche} and \emph{s-trans} conformers, we selected the data
points at the locations labeled $A$ and $B$ in the figure. From our simulations, we estimate that
the populations are $p_\mathit{s-trans}\approx1$ at $A$ and $p_\mathit{s-trans}\approx0$ at $B$. We
evaluate the imbalance between \emph{gauche} and \emph{s-trans} for any fragment $X$ as the relative
difference $a_X=(n_{X}^{A}-n_X^{B})/(n_{X}^{A}+n_X^{B})$ with
$n_{X}^{A,B}=N_{X}^{A,B}/N_\mathrm{VUV}^{A,B}$ being the fragment counts $N_{X}^{A,B}$ normalized by
the total DBB beam density $N_\mathrm{VUV}^{A,B}$ at the respective point as measured by VUV
ionization. The imbalance $a_X$ takes values in the range $[-1,+1]$, corresponding to a strong
correlation with \emph{s-trans} or \emph{gauche} DBB, respectively. \autoref{fig:imbalance}~b) shows
the obtained imbalance values which range from $-0.9(1)$ for DBB$^+$ to $0.60(7)$ for Br$^+$. All
fragments show a tendency of increasing imbalance towards $\emph{gauche}$ with decreasing fragment
size, thus suggesting that \emph{gauche} DBB is more likely to break up into smaller parts during
the interaction with the fs laser pulse. A rationalization of this phenomenon requires further
study.

\section{Conclusions}
\label{sec:conclusion}
Driven by the motivation to gain a precise understanding of the effects of molecular conformation in
cycloaddition reactions, a quantum-chemical screening was performed to identify diene
candidates suitable for conformer separation in a molecular-beam apparatus. As an optimal diene,
2,3-dibromobuta-1,3-diene was found to exhibit the desired large difference in electric dipole
moments and small energy difference between the two conformers. Since this particular dihalogenated
diene cannot be purchased, mainly due to its intrinsic tendency to undergo polymerization, a
synthesis was developed to produce the compound in adequate purity. Experimental validation of the
calculated properties was achieved by seeding DBB in a molecular beam and separating its
\emph{gauche} and \emph{s-trans} conformers in an electrostatic field gradient. A deflection profile
of DBB was measured by subsequent ionization and ejection into a time-of-flight mass-spectrometer.
The implementation of a vacuum-ultraviolet light source achieved ionization of the parent diene
without fragmentation and therefore made it possible to directly measure its deflection behavior.
Spatial separation of the two conformers was then confirmed by a close agreement of the observed
deflection profiles with Monte-Carlo simulations based on the theoretical molecular properties.
Comparison between ion yields from VUV and non-resonant fs-laser-pulse ionization suggest different
fragmentation patterns for the \emph{gauche} and \emph{s-trans} conformers during ionization in the
strong field. The polar \emph{gauche} conformer showed an enhanced tendency to fragment in
comparison with the apolar \emph{s-trans} conformer. The successful separation of the \emph{gauche}
and \emph{s-trans} conformers of this tailor-made diene paves the way toward studies of
conformer-selected polar cycloaddition reactions in a cold and controlled environment.

\section*{Conflicts of interest}
The authors declare no competing financial or non-financial interests.

\section*{Acknowledgements}
We thank Philipp Kn\"opfel, Grischa Martin and Georg Holderried for technical support. Marco Meyer and
Jia Wang are acknowledged for their assistance with the experiments. We thank Max Schwilk for
fruitful discussions on the theoretical results. This work is supported by the Swiss National
Science Foundation (Nr.~BSCGI0\_157874). O.A.v.L.\ acknowledges further funding from the Swiss
National Science foundation (Nr.~PP00P2\_138932 and 407540\_167186 NFP 75 Big Data) and from the
European Research Council (ERC-CoG grant QML). This work was partly supported by the NCCR MARVEL,
funded by the Swiss National Science Foundation. H.G.\ and S.W.\ acknowledge support by the K.C.\
Wong Education Foundation.

\bibliography{all_refs}%Main-Oct17,newrefs_dbb}

\clearpage
\onecolumngrid

\begin{center}
{\bf \large Supplementary Information}
\end{center}
\setcounter{section}{0}
\setcounter{figure}{0}
\setcounter{table}{0}
\setcounter{equation}{0}

\renewcommand\thesection{S\arabic{section}}
\renewcommand\thefigure{S\arabic{figure}}
\renewcommand\thetable{S\arabic{table}}
\renewcommand\theequation{S\arabic{equation}}

\section{Synthesis of 2,3-dibromobutadiene}
\label{sec:SI:synthesis}
\subsection*{Materials and methods} % used chemicals and NMR spectrometer
\begin{sidewaysfigure}[htbp]
   \centering
   \includegraphics[width=0.8\linewidth]{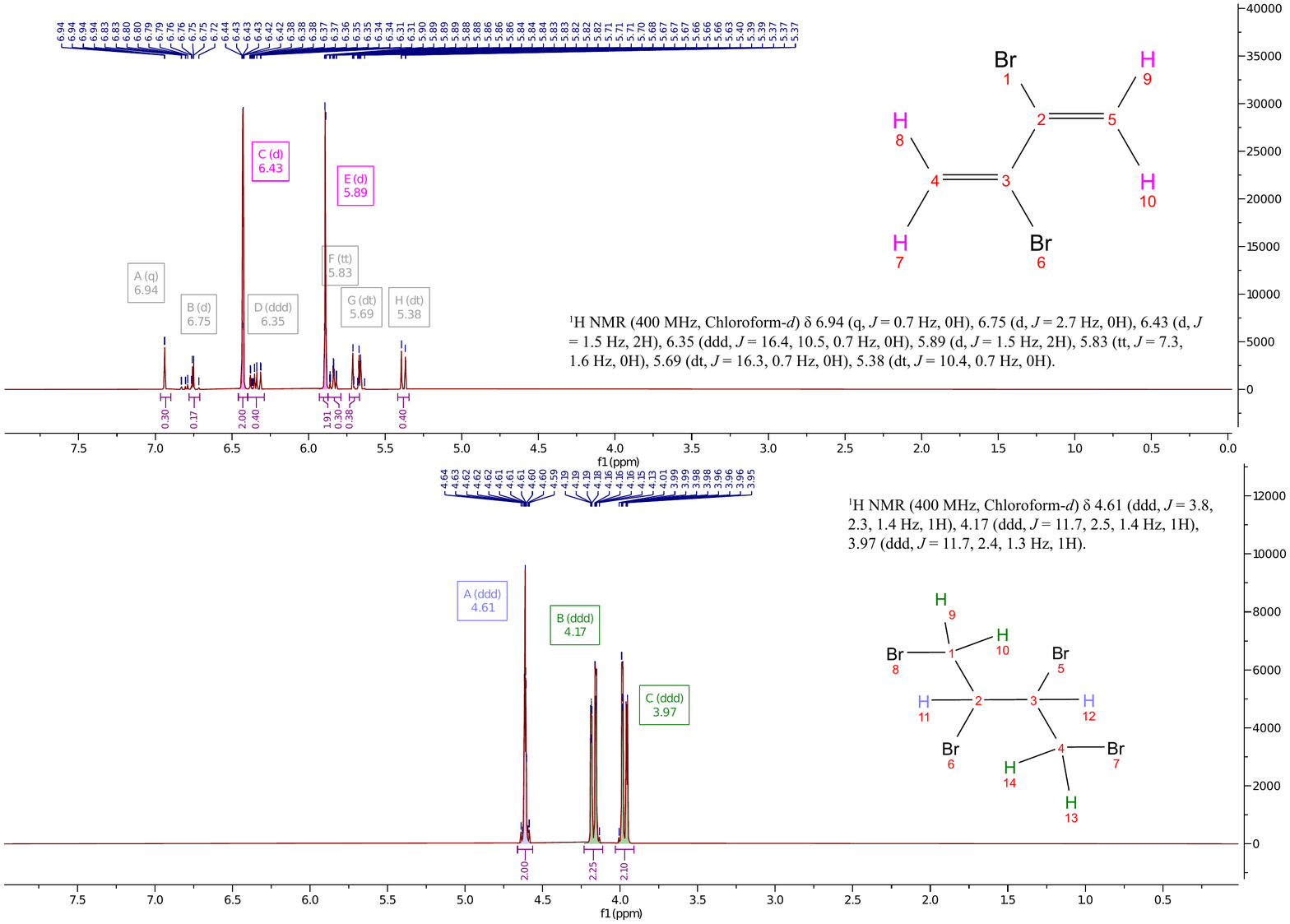}
   \caption{\textbf{Analysis.} $^1$H NMR analysis of the reactant 1,2,3,4-tetrabromobutane (bottom)
      and the product 2,3-dibromobuta-1,3-diene (top).}
   \label{fig:NMR}
\end{sidewaysfigure}

Diethyl ether and sodium iodide were purchased from Sigma-Aldrich.
\mbox{1,8-Diazabicyclo[5.4.0]undec-7-ene} was purchased from Alfa Aesar. 1,2,3,4-tetrabromobutane
was purchased from TCI-chemicals. All reagents were used without further purification. Chloroform
for NMR measurements was purchased from Cambridge Isotope Laboratories. All $^{1}$H and $^{13}$C NMR
spectra (Fig.~\ref{fig:NMR})were recorded on a Bruker Avance III (HD)NMR instrument operated at 400 MHz and 101 MHz,
respectively. Chemical shifts ($\delta$) are reported in parts per million (ppm) relative to residual solvent peaks.

\subsection*{Synthesis of 2,3-dibromobutadiene} % Synthesis od DBB

\begin{figure}[h!]
   \centering%
   \includegraphics[width=0.8\linewidth]{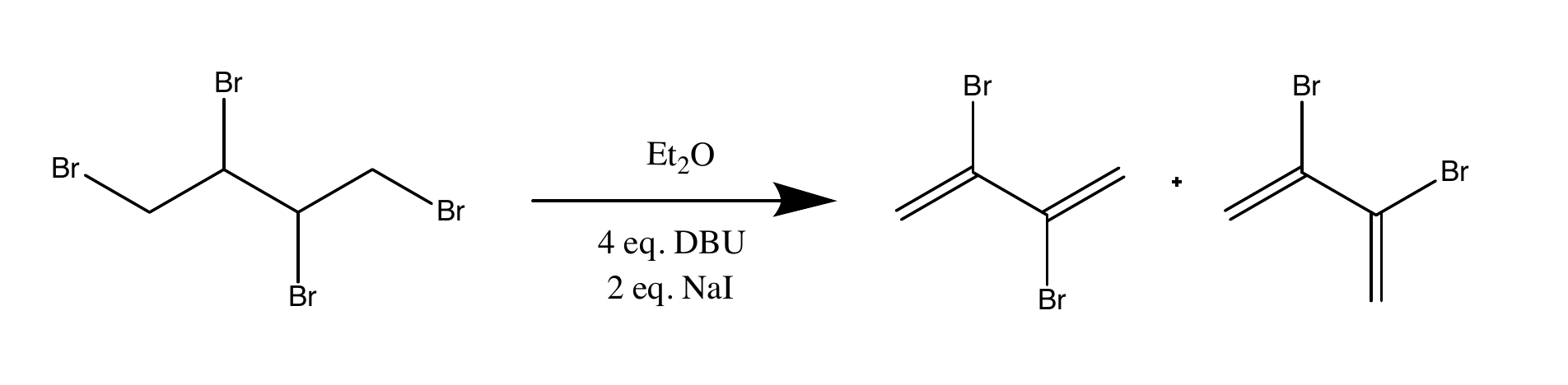}
   \caption{\textbf{Overview of the synthesis of 2,3-dibromobutadiene.}}
   \label{fig:synthesis}
\end{figure}

An overview of the synthesis of 2,3-dibromobutadiene is shown in Fig.~\ref{fig:synthesis}.

In a 500~ml three-neck round bottom flask 1,2,3,4-tetrabromobutane (12.0~g, 32.12~mmol, 1~eq) were
suspended in diethyl ether (150~ml, extra pure, stabilized with BHT). Sodium iodide (9.64~g,
64.24~mmol, 2~eq., anhydrous $>99.5~\%$) was added to the suspension. A flow of nitrogen was
continuously passed through the flask during the entire operation.

By making use of a dropping funnel, 1,8-Diazabicyclo[5.4.0]undec-7-ene (19.2~ml, 1.29~mol, 4~eq.)
was slowly added to the suspension. During the addition the formation of a dense yellow suspension
was observed. The dropping funnel was rinsed with diethyl ether (30~ml) and the reaction mixture was
allowed to stir for 1~hour. Then, the precipitate was filtered off over a frit and was washed with
diethyl ether ($3\times20$~ml). The organic phase was washed with saturated ammonium chloride
($3\times100$~ml), distilled water ($1\times100$~ml), and brine ($1\times100$~ml). The organic phase
was dried with MgSO$_4$ and concentrated using a rotatory evaporator at 0~\textdegree{C} and reduced
pressure to yield the product DBB (3.74~g, $55~\%$). The product was observed to react violently
with excessive heat formation when in contact with air at room temperature. In order to prevent the
sample from degradation and self-polymerization it was stored below $-78$~\textdegree{C}.

\begin{description}
\item[$^{1}$H NMR (400 MHz, CDCI3)]
$\delta$ 6.43 (d, $J=1.5$~Hz, 2H), 5.89 (d, $J=1.5$~Hz, 2H)

\item[$^{13}$C NMR (101 MHz, CDCI3)]
$\delta$ 125.13, 124.75

\item[LRMS (EI)]
calculated for \ch{C_{4}H_{4}Br_{2}}: 211.87, found: 211.85
\end{description}

% NMR analysis

%\clearpage
%\section{Additional theoretical results}
%\label{sec:SI:theorsup}
%\begin{figure}
%    \centering
%    \includegraphics[scale=0.35]{SI_fig_1}
%    \caption{Potential energy surface of dibromo-butadiene (CH$_2$=C(Br)-C(Br)=CH$_2$), dichloro-butadiene (CH$_2$=C(Cl)-C(Cl)=CH$_2$), and difluoro-butadiene (CH$_2$=C(F)-C(F)=CH$_2$) (eV) calculated using DSD-PBEP86-D3BJ. Due to the symmetry of the molecules, the torsional profiles are symmetric with respect to mirroring at $\Theta=180\,^\circ$.}
%    \label{fig:torsional_profile_br_cl_f}
%\end{figure}

\end{document}